\documentclass[11pt,a4paper,english]{article}

\usepackage{graphicx}
\usepackage{amssymb,latexsym}
\usepackage{amsmath}
\usepackage{epsf}
\usepackage{cite}
\usepackage{pdfsync}
\usepackage{epsfig}
\usepackage{wrapfig}
\usepackage[matrix,arrow,color]{xy}
\usepackage[usenames]{color}
\usepackage{hyperref}
\usepackage{bbm}
\usepackage{comment}
\usepackage[font=footnotesize,labelfont=bf,margin=1cm]{caption}

\input{xy}
\xyoption{all}

\input{epsf}

\makeatletter

\catcode`@=11

\numberwithin{equation}{section}
\catcode`@=12

\newcommand{\ba}{\begin{eqnarray*}}
\newcommand{\ea}{\end{eqnarray*}}
\newcommand{\ban}{\begin{eqnarray}}
\newcommand{\ean}{\end{eqnarray}}

\newcommand{\IZ}{\mathbb{Z}}

\newcommand{\IR}{\mathbb{R}}

\newcommand{\mbf}[1]{{\boldsymbol {#1} }}

\def\ii{{\,{\rm i}\,}}
\def\dd{{\rm d}}

\def\beq{\begin{equation}}
\def\bee{\begin{equation}}
\def\eeq{\end{equation}}
\def\bea{\begin{eqnarray}}
\def\eea{\end{eqnarray}}
\def\bd{\begin{displaymath}}
\def\ed{\end{displaymath}}

\newcommand{\Cint}{\int\kern-10.5pt-\kern7pt}

\newcommand{\be}{\begin{equation}}
\newcommand{\ee}{\end{equation}}

\newcommand\fverbit{\egroup\item[\fbox{\unhbox\pippobox}]}
\newbox\pippobox

{

\def\be{\begin{equation}}
\def\ee{\end{equation}}
\def\bea{\begin{eqnarray}}
\def\eea{\end{eqnarray}}

\begin{document}

\begin{titlepage}
\setcounter{page}{0}
\begin{flushright}
LMU--ASC 33/17 \\
MPP--2017--106\\
EMPG--17--07
\end{flushright}

\vskip 1.5cm

\begin{center}

{\Large\bf Non-geometric Kaluza-Klein monopoles \\[2mm] and magnetic
  duals of M-theory $\mbf R$-flux backgrounds}

\vspace{15mm}

{\large\bf Dieter L\"ust$^{(a),(b)}$} \ , \ {\large\bf Emanuel Malek$^{(a)}$} \ and \ 
{\large\bf Richard~J.~Szabo$^{(c)}$}
\\[6mm]

\noindent{\em $^{(a)}$ Arnold Sommerfeld Center for Theoretical
  Physics, Department f\"ur Physik\\ Ludwig-Maximilians-Universit\"at
  M\"unchen\\ Theresienstra{\ss}e 37, 80333 M\"unchen, Germany} \\ Email: \ {\tt
  dieter.luest@lmu.de \ , \ e.malek@lmu.de}\\[4mm]
\noindent{\em $^{(b)}$ Max-Planck-Institut f\"ur Physik,
  Werner-Heisenberg-Institut\\ F\"ohringer Ring 6, 80805 M\"unchen, Germany
}\\[4mm]
\noindent{\em $^{(c)}$ Department of Mathematics\\ Heriot-Watt
  University\\
Colin Maclaurin Building, Riccarton, Edinburgh EH14 4AS, UK\\ 
Maxwell Institute for Mathematical Sciences, Edinburgh, UK\\
The Higgs Centre for Theoretical Physics, Edinburgh, UK}\\
Email: \ {\tt
  R.J.Szabo@hw.ac.uk}

\vspace{20mm}

\begin{abstract}
\noindent

We introduce a magnetic analogue of the seven-dimensional nonassociative octonionic $R$-flux algebra that describes the phase space of M2-branes in four-dimensional locally non-geometric M-theory backgrounds. We show that these two algebras are related by a $Spin(7)$ automorphism of the 3-algebra that provides a covariant description of the eight-dimensional M-theory phase space. We argue that this algebra also underlies the phase space of electrons probing a smeared magnetic monopole in quantum gravity by showing that upon appropriate contractions, the algebra reduces to the noncommutative algebra of a spin foam model of three-dimensional quantum gravity, or to the nonassociative algebra of electrons in a background of uniform magnetic charge. We realise this set-up in M-theory as M-waves probing a delocalised Kaluza-Klein monopole, and show that this system also has a seven-dimensional phase space. We suggest that the smeared Kaluza-Klein monopole is non-geometric because it cannot be described by a local metric. This is the magnetic analogue of the local non-geometry of the $R$-flux background and arises because the smeared Kaluza-Klein monopole is described by a $U(1)$-gerbe rather than a $U(1)$-fibration.

\end{abstract}

\end{center}


\end{titlepage}

\newpage


\tableofcontents

\bigskip

\section{Introduction and summary} \label{intro}

It is well-known that the effective theory of open strings on D-branes with flux is a noncommutative gauge theory. However, it has long been unclear whether noncommutative geometry has any concrete realisation within a similar context in the closed string sector of string theory, whereby it can provide an effective target space description for quantum gravity. In recent years it has been found that the closed string sector in a non-geometric background, where the usual geometric notions of target space break down, does reveal noncommutative and even nonassociative deformations of spacetime geometry~\cite{Blumenhagen:2010hj,Lust:2010iy,Blumenhagen:2011ph,Condeescu:2012sp,Plauschinn:2012kd,Mylonas:2012pg,Andriot:2012vb,Blair:2014kla,Bakas:2015gia}, which can be quantised in the framework of deformation quantisation, see e.g. \cite{Mylonas:2012pg}.

In this context `non-geometry' refers to the feature that one cannot define the metric and
$p$-form fluxes of supergravity globally -- such backgrounds are
termed ``globally non-geometric'' spaces -- or even locally in the
case of ``locally non-geometric'' spaces. The measure of global and
local non-geometry is provided by non-geometric $Q$-flux and
$R$-flux, respectively. At the same time, these also parameterise
noncommutativity and nonassociativity of the space, with the
noncommutativity related to both $Q$-flux and $R$-flux. On the
other hand, spatial nonassociativity is parameterised by the $R$-flux alone, and thus only occurs in locally non-geometric backgrounds.

In this paper we will study three instances of nonassociative
deformations and the relations between them. The first is given by the
example of closed strings propagating in a constant $R$-flux background. In this case the
nonassociative coordinate algebra of the closed string phase space is given by
\begin{equation}
 \left[ x^i, x^j \right] = \frac{\ii\ell_s^3}{\hbar} \, R^{ijk} \, p_k
 \ , \qquad \left[ x^i, p_j \right] = \ii \hbar \, \delta^i_j \ ,
 \qquad \big[ p_i, p_j \big] = 0 \ , \label{eq:StringRFluxAlgebra}
\end{equation}
which has non-vanishing Jacobiator amongst the position coordinates given by
\begin{equation}
[\![x^i,x^j,x^k]\!] := \mbox{$\frac13$} \, \big( [ [x^i, x^j ], x^k ]
+ [ [ x^j, x^k ], x^i ] + [ [ x^k, x^i ], x^j ] \big) = \ell_s^3 \,
R^{ijk} \ .
\end{equation}
As pointed out by~\cite{Mylonas:2012pg}, the
violation of associativity can be attributed geometrically to a noncommutative gerbe with connective structure on phase space whose
curvature three-form has components given by the flux $R^{ijk}$
along the momentum space directions; alternatively, we may regard it as a gerbe on
the doubled space of double field theory with curvature along the ``dual
coordinates'' in a suitable solution to the section constraints. This coordinate algebra, like the others we will be encountering in this paper, can be quantised~\cite{Mylonas:2012pg} despite its nonassociative nature.

The second nonassociative algebra we shall study is that governing the phase space of an electron in three spatial dimensions in the presence of magnetic charges, which is given by~\cite{Jackiw:1984rd,Grossman:1984fs,Wu:1984wr,Jackiw:1985hq,Mickelsson:1985fa,Gunaydin:1985ur,Bakas:2013jwa}
\begin{equation}
 \left[ x^i, x^j \right] = 0 \,, \qquad \left[ x^i , p_j \right] =
 \ii\hbar \, \delta^i_j \,, \qquad \big[ p_i, p_j \big] = \ii 
 \hbar \, \varepsilon_{ijk} \, B^k \,, \label{eq:ElectronBFluxAlgebra}
\end{equation}
where $\vec B$ is the magnetic field, $x^i$ are the electron
coordinates and $p_i$ are the \emph{gauge-invariant} momenta. There
is now a non-trivial Jacobiator amongst the momentum coordinates given by
\begin{equation}
 [\![ p_i, p_j, p_k]\!] = - \hbar^2 \, \varepsilon_{ijk} \, \nabla \cdot \vec B \,,
\end{equation}
where $\nabla \cdot \vec B = \partial_i B^i = \rho_{\rm mag}$ is the magnetic charge density. Thus the coordinate algebra violates associativity precisely at the loci of the magnetic charges. In the case of isolated magnetic charges, one can avoid the nonassociativity by excising these points in position space. By angular momentum conservation, the electrons never reach the magnetic monopoles~\cite{Bakas:2013jwa}, and the electron wavefunctions necessarily vanish at their locations~\cite{Jackiw:1984rd}, motivating their removal from position space. This excision is of course also motivated from the fact that the magnetic field itself and its associated gauge field are singular at the loci of the magnetic sources, and is commonly used in the geometric description of monopoles as connections on non-trivial fibre bundles over the excised space.

Now consider a constant magnetic charge density $\rho_{\rm mag} =
N / \ell_m^3$, where $\ell_m$ is the characteristic magnetic scale. Because the magnetic charge is now uniformly distributed throughout all of space, the coordinate algebra
\eqref{eq:ElectronBFluxAlgebra} becomes everywhere nonassociative; if
one wanted to remove the magnetic sources from position space, one would be
left with empty space. This signals that the magnetic charge
example also shows signs of local non-geometry, just like the $R$-flux
example does. This analogy can in fact be made more precise: The
algebra \eqref{eq:ElectronBFluxAlgebra} is isomorphic to the
nonassociative string theory $R$-flux algebra \eqref{eq:StringRFluxAlgebra} in three spatial dimensions with the isomorphism given schematically by
\begin{equation}
 \begin{split}
  \frac{1}{\ell_s}\,x_R^i &\longleftrightarrow \frac{\ell_m}{\hbar}\, p_{M\,i} \ , \\[4pt]
  \ell_s \, p_{R\,i} &\longleftrightarrow \frac{\hbar}{\ell_m}\, x_M^i \ , \\[4pt]
  R = \mbox{$\frac1{3!}$} \, \varepsilon_{ijk} \, R^{ijk}
  &\longleftrightarrow \frac{N}{\hbar} \, \label{eq:RMagIso}
 \end{split}
\end{equation}
where we have included the subscripts $R / M$ to distinguish the phase space variables of the $R$-flux/magnetic backgrounds, respectively.
In this case, just as for the constant $R$-flux example, the
nonassociativity cannot be avoided. Furthermore, the constant magnetic
charge density can only be described geometrically in terms of a connective structure
on a gerbe rather than a connection on a fibre bundle, again similarly to the $R$-flux case, although in this
instance the gerbe resides over position space.

The third nonassociative algebra that we consider here is given by the
seven imaginary units of the octonions. Recent work~\cite{Gunaydin:2016axc}
on non-geometric backgrounds in M-theory has suggested that these
underlie the lift of the string theory $R$-flux algebra to
M-theory. Crucially, this identification relies on the fact that in
four-dimensional locally non-geometric M-theory backgrounds there is a missing momentum mode and hence the phase space is seven-dimensional. The missing momentum constraint was conjectured to take the form
\begin{equation}
 R^{\mu,\nu\rho\alpha\beta} \, p_\mu = 0 \ , \label{eq:MomentumConstraint}
\end{equation}
where $\mu, \nu,\ldots = 1, \ldots, 4$ and $R^{\mu,\nu\rho\alpha\beta}$ is the M-theory $R$-flux \cite{Blair:2014zba} which parameterises the local non-geometry. As discussed in~\cite{Gunaydin:2016axc}, the constraint \eqref{eq:MomentumConstraint} can be understood by duality to the twisted torus, wherein a particular two-cycle is homologically trivial, or equivalently by a further duality to the Freed-Witten anomaly. In the string theory limit it implies that there are no D0-branes in the $R$-flux background, which has also been argued before~\cite{Wecht:2007wu}, and which further agrees with expectations from quantum mechanics on the nonassociative space~\cite{Mylonas:2013jha}. These observations, together with the fact that the octonion algebra can be contracted to the string theory $R$-flux algebra \eqref{eq:StringRFluxAlgebra}, led to the conjecture~\cite{Gunaydin:2016axc} that the M-theory $R$-flux algebra is to be identified with the octonion algebra and hence is given by
\small
\begin{equation}
\begin{split}
\left[x^i,x^j\right] =&\, \frac{\ii\ell_s^3}{\hbar}\, R^{4,ijk4}\, p_k \,, \qquad
\left[x^4,x^i\right] = \frac{\ii \lambda\, \ell_s^3}{\hbar} R^{4,1234} p^i \,, \\[4pt]
\left[x^i,p_j\right] =&\, \frac{\ii\hbar}{\ell_s}\,\delta^i_j\,x^4+\frac{\ii\hbar\,\lambda}{\ell_s}\, \varepsilon^i{}_{jk}\, x^k \,,
\qquad \left[p_i,x^4\right] = \frac{\ii\hbar\,\lambda^2}{\ell_s} \,x_i \,, \\[4pt]
\big[p_i,p_j\big] =&\, -\frac{\ii\hbar\,\lambda}{\ell_s}\, \varepsilon_{ijk}\, p^k\,, \\[4pt]
[\![x^i,x^j,x^k]\!] =& \, -\ell_s^2 \, R^{4,ijk4} \, x^4 \,, \qquad 
[\![x^i,x^j,x^4]\!] = \, \lambda^2\, \ell_s^2\, R^{4,ijk4} \, x_k \,, \\[4pt]
[\![p_i,x^j,x^k]\!] =& \, -\lambda\,\ell_s^2 \, R^{4,1234}\, \big(\delta^j_i\, p^k-\delta^k_i\, p^j \big) \,, \\
[\![p_i,x^j,x^4]\!] =& \, -\lambda^2\, \ell_s^2\, R^{4,ijk4}\, p_k \,, \\[4pt]
[\![p_i,p_j,x^k]\!] =& \, \frac{\hbar^2\, \lambda}{\ell_s^2}\, \varepsilon_{ij}{}^{k}\, x^4 + \frac{\hbar^2\, \lambda^2}{\ell_s^2}\, \big(\delta_j^k\, x_i-\delta_i^k\, x_j \big) \,, \\[4pt]
[\![p_i,p_j,x^4]\!] =& \, -\frac{\hbar^2\, \lambda^3}{\ell_s^2}\, \varepsilon_{ijk}\, x^k \,, \qquad
[\![p_i,p_j,p_k]\!] = \, 0 \,, \label{eq:MRFluxAlgebra}
\end{split}
\end{equation}
\normalsize
where $\lambda$ is the dimensionless contraction parameter.

In~\cite{Kupriyanov:2017oob} it was shown that the M-theory $R$-flux
algebra, like the other examples considered here, can also be
quantised based on the observation that it is determined by a choice of
$G_2$-structure on the seven-dimensional phase space. In turn, the
$G_2$-structure can be naturally extended to a $Spin(7)$-structure
in eight dimensions, suggesting an algebraic framework for
understanding the full unconstrained M-theory phase space in terms of
the full eight-dimensional algebra of octonions. Based on this observation, and the general expectations that higher 3-algebra structures underlie the lift of string theory to M-theory, in~\cite{Kupriyanov:2017oob} it was proposed that the full eight-dimensional $Spin(7)$-covariant phase space is described by lifting \eqref{eq:MRFluxAlgebra} to a 3-algebra with 3-bracket relations
\small
\begin{equation}\label{eq:3algebraintro}
\begin{split}
[x^i,x^j,x^k] &=\,- \frac{\ell_s^2}{2}  \, R^{4,ijk4}\, x^4 \,,\qquad
[x^i,x^j,x^4] =  \frac{\lambda^2\,\ell_s^2}{2} \, R^{4,ijk4}\, x_k \,,  \\[4pt]
[p^i,x^j,x^k] &=\,- \frac{\lambda\,\ell_s^2}{2} \, R^{4,ijk4}\, p_4- \frac{\lambda\,\ell_s^2}{2} \, R^{4,1234}\, \big(\delta^{ij}\, p^k - \delta^{ik}\, p^j\big) \,,\\[4pt]
[p_i,x^j,x^4] &=\,- \frac{\lambda^2\,\ell_s^2}{2} \, R^{4,1234} \, \delta_i^j\, p_4- \frac{\lambda^2\,\ell_s^2}{2} \, R^{4,ijk4}\, p_k \,, \\[4pt]
[p_i,p_j,x^k] &=\,  \frac{\hbar^2\,\lambda}{2\,\ell_s^2} \, \varepsilon_{ij}{}^k\, x^4+ \frac{\hbar^2\,\lambda^2}{2\,\ell_s^2} \,\big(\delta_j^k\, x_i-\delta_i^k\, x_j\big) \,, \\[4pt]
[p_i,p_j,x^4] &=\, - \frac{\hbar^2\,\lambda^3}{2\,\ell_s^2} \,\varepsilon_{ijk}\, x^k \,, \qquad
[p_i,p_j,p_k] = -2\,\frac{\hbar^2\,\lambda^2}{\ell_s^2} \,\varepsilon_{ijk}\, p_4 \,, \\[4pt]
[p_4,x^i,x^j] &=\, \frac{\lambda\, \ell_s^2}{2} \, R^{4,ijk4}\, p_k \,, \qquad
[p_4,x^i,x^4] = - \frac{\lambda^2\, \ell_s^2}{2} \, R^{4,1234} \, p^i \,, \\[4pt]
[p_4,p_i,x^j] &=\, - \frac{\hbar^2\,\lambda}{2\,\ell_s^2} \, \delta_i^j\, x^4 -  \frac{\hbar^2\,\lambda^2}{2\ell_s^2}  \, \varepsilon_i{}^{jk}\, x_k \,, \\[4pt]
[p_4,p_i,x^4] &=\, - \frac{\hbar^2\,\lambda^3}{2\, \ell_s^2}  \, x_i \,, \qquad
[p_4,p_i,p_j] = - \frac{\hbar^2\,\lambda^2}{2\, \ell_s}  \, \varepsilon_{ijk}\, p^k \,.
\end{split}
\end{equation}
\normalsize
The 3-algebra \eqref{eq:3algebraintro} in turn reduces to the algebra
\eqref{eq:MRFluxAlgebra} when the momentum constraint
\eqref{eq:MomentumConstraint} is implemented as a type of gauge-fixing constraint.

In this paper we shall begin by showing that the isomorphism \eqref{eq:RMagIso} can be understood as an automorphism of the $Spin(7)$ symmetry group underlying the 3-algebra~\eqref{eq:3algebraintro}. 
The automorphism exchanges the $R$-flux phase space with that of the magnetic monopole, and correspondingly a missing momentum coordinate for a missing position coordinate. Similarly, the gerbe over momentum space, the $R$-flux, becomes a gerbe over position space, the smeared magnetic monopole. This suggests that the $R$-flux and magnetic monopole backgrounds are related by a sort of ``canonical transformation'', preserving the 3-algebra \eqref{eq:3algebraintro}. 

Given this observation, it is natural to wonder what role the octonion algebra plays in the case of the three-dimensional magnetic monopole. We will argue that this seven-dimensional phase space is related to a magnetic monopole in quantum gravity since it involves two parameters, $\lambda$ and $N$ (when the M-theory $R$-flux is given by $R^{4,\mu\nu\rho\sigma} = N \, \varepsilon^{\mu\nu\rho\sigma}$), whose contraction limits respectively lead to the magnetic monopole algebra \eqref{eq:ElectronBFluxAlgebra} and the noncommutative algebra underlying a certain Ponzano-Regge model of three-dimensional quantum gravity~\cite{Freidel:2005me}. This suggests that the parameters we identify are the Planck length $\ell_{\rm P}$ and the magnetic charge.

To explicitly realise this picture and to further understand the missing phase space coordinate constraint,
we also embed a magnetic monopole into string theory
and M-theory where it can be realised as a D6-brane and its
Kaluza-Klein monopole uplift. The electron probes of this
background are realised as D0-branes and M-waves, respectively. From
this perspective we obtain a natural interpretation of the seventh
phase space coordinate in the M-theory lift as the momentum along the
Taub-NUT circle. 

While the usual
Kaluza-Klein monopole solution is based on Dirac monopoles, in order to accomodate for the uniform distribution
of magnetic charge, or equivalently of D6-brane sources, we actually need a non-local variant defined by smearing the standard solution; its
metric has no local expression, which is an earmark of local non-geometry
and hence we call this solution a ``non-geometric Kaluza-Klein
monopole'', being based on a gerbe rather than a fibre bundle
as in the standard case. Then the canonical transformation 
above provides an isomorphism between the phase spaces of M2-branes in
the non-geometric $R$-flux background of M-theory and of M-waves in
the non-geometric Kaluza-Klein monopole background of M-theory, which are both
provided by the nonassociative algebra of the seven imaginary unit
octonions. The contraction limit $\lambda\to0$ turns off quantum
gravitational effects, i.e. the effects of finite string coupling with
$g_s,\ell_{\rm P} \rightarrow 0$, and describes the propagation
of a string in the $R$-flux background or of an electron in the field
of a constant magnetic charge density.

The outline of the remainder of this paper is as follows. We begin by reviewing the M-theory $R$-flux algebra in \S\ref{s:MRFluxReview} and then describe its magnetic dual via a $Spin(7)$ automorphism of the 3-algebra \eqref{eq:3algebraintro} in \S\ref{sec:magneticdual}, that we show can be reduced to the phase space algebra underlying a certain spin foam model of three-dimensional quantum gravity in \S\ref{sec:mongravity}. In \S\ref{sec:KKmon} we describe this magnetic dual by a delocalised Kaluza-Klein monopole in M-theory, which does not admit a local metric. We introduce M-wave probes of this background in \S\ref{s:MwavePSpace}, and argue that they only have a seven-dimensional phase space, similarly to M2-branes in the $R$-flux background. Finally, in \S\ref{s:8dPSpace} we show that in the limit where there is no magnetic charge and no $R$-flux, the two eight-dimensional $Spin(7)$-symmetric 3-algebras become completely identical and should describe the unique non-associative eight-dimensional phase space of M-theory, possibly even in the decompactification limit of the M-theory circle.

\section{M2-brane phase space and its magnetic dual \label{sec:M2}}

\subsection{Nonassociative M-theory $R$-flux background and M2-branes} \label{s:MRFluxReview}

In locally non-geometric M-theory in four dimensions with constant $R$-flux, the eight-dimensional phase space is described by a 3-algebra structure which is invariant under a subgroup $Spin(7)\subseteq SO(8)$~\cite{Kupriyanov:2017oob}. The brackets of this 3-algebra are given by
\bea
\big[{\mit\Xi}_{\hat A},{\mit\Xi}_{\hat B},{\mit\Xi}_{\hat C}\big] = -2\,\hbar^2\,  \phi_{\hat A\hat B\hat C\hat D}\, {\mit\Xi}_{\hat D} \ ,
\label{eq:3algebrageneral}\eea
where $({\mit\Xi}_{\hat A})$ parameterise $\IR^8$ and the rank four
antisymmetric tensor $\phi_{\hat A\hat B\hat C\hat D}$ is related to
the structure constants of the eight-dimensional nonassociative
algebra of octonions. The failure of \eqref{eq:3algebrageneral} in
defining a Nambu-Poisson bracket is measured by its violation of the
fundamental identity through the 5-bracket
\beq
\begin{split} \label{eq:HigherNonAssoc}
\big[\!\!\big[{\mit\Xi}_{\hat A},{\mit\Xi}_{\hat B},{\mit\Xi}_{\hat
  C},{\mit\Xi}_{\hat D},{\mit\Xi}_{\hat
  E}\big]\!\!\big]&:=\mbox{$\frac1{12}$}\, \big(\,
\big[ \big[ {\mit\Xi}_{\hat A},{\mit\Xi}_{\hat B},{\mit\Xi}_{\hat
  C} \big] ,{\mit\Xi}_{\hat E},{\mit\Xi}_{\hat
  D} \big] +
\big[ \big[ {\mit\Xi}_{\hat A},{\mit\Xi}_{\hat B},{\mit\Xi}_{\hat
  D}\big] ,{\mit\Xi}_{\hat C},{\mit\Xi}_{\hat
  E}\big] \\ & \qquad\qquad +
\big[ \big[ {\mit\Xi}_{\hat A},{\mit\Xi}_{\hat B},{\mit\Xi}_{\hat
  E}\big] ,{\mit\Xi}_{\hat C},{\mit\Xi}_{\hat
  D}\big] + 
\big[ \big[ {\mit\Xi}_{\hat
  C},{\mit\Xi}_{\hat D},{\mit\Xi}_{\hat E}\big] ,
{\mit\Xi}_{\hat A},{\mit\Xi}_{\hat B} \big] \, \big)\\[4pt]
&=\,
\hbar^4\, \big(\,\delta_{\hat A\hat C}\,\phi_{\hat D\hat E\hat B\hat
  F}+\delta_{\hat A\hat D}\,\phi_{\hat E\hat C\hat B\hat
  F}+\delta_{\hat A\hat E}\,\phi_{\hat C\hat D\hat B\hat F} \\ & 
\qquad \qquad -\, \delta_{\hat B\hat C}\,\phi_{\hat D\hat E\hat A\hat F} 
-\delta_{\hat B\hat D}\,\phi_{\hat E\hat C\hat A\hat F}
-\delta_{\hat B\hat E}\,\phi_{\hat C\hat D\hat A\hat
  F}\,\big)\,{\mit\Xi}^{\hat F}\\ &\qquad
-\,\hbar^4\, \big( \,\phi_{\hat B\hat C\hat D\hat
  E}\, {\mit\Xi}_{\hat A}-\phi_{\hat A\hat C\hat D\hat E}\,{\mit\Xi}_{\hat B}\,\big) \ .
\end{split}
\eeq
The fundamental identity plays the higher 3-algebraic analogue of the Jacobi identity and thus the failure of the 5-bracket \eqref{eq:HigherNonAssoc} to vanish represents a ``higher non-associativity'', as expected in the M-theory lift of string theory. This also matches the fact that while the $R$-flux in string theory is represented by a rank 3 tensor, its M-theory uplift is a rank 5 tensor.

Consider any constraint
\bea
f({\mit\Xi})=0 \,,
\eea
which reduces the phase space by one dimension and breaks the symmetry
as $Spin(7) \longrightarrow G_2$. Then the action of the group
$Spin(7)$ on the spinor representation $\mbf 8$ restricts to $G_2$
according to the decomposition
\begin{equation}
 {\bf 8}\big|_{G_2} = \mbf{7} \oplus \mbf{1} \ ,
\end{equation}
where $f({\mit\Xi})$ lives in the singlet representation and the remaining coordinates, which we denote by $(\xi_A)$, live in the seven-dimensional representation. The resulting seven-dimensional phase space is governed by a nonassociative algebra invariant under $G_2\subseteq SO(7)$ with brackets
\bea
[\xi_A,\xi_B] := [\xi_A,\xi_B,f] \ ,
\eea
whose Jacobiators $[\![\xi_A,\xi_B,\xi_C]\!]$ are obtained by setting $f({\mit\Xi})=0$ in the
remaining 3-algebra relations $[\xi_A,\xi_B,\xi_C]$ from \eqref{eq:3algebrageneral}.
For the particular choice of constraint function
$f({\mit\Xi})={\mit\Xi}_8$, which arises from turning on the
$R$-flux~\cite{Gunaydin:2016axc}, these are just the commutation relations of the nonassociative octonion algebra~\cite{Kupriyanov:2017oob}.

The parameterisation $({\mit\Xi}_{\hat A})\longrightarrow (x^\mu,p_\mu)\in\IR^4\times\IR^4$ of the eight-dimensional phase space into position and momentum coordinates further breaks the symmetry group to $SU(2)^3 \subseteq Spin(7)$. The presence of $R$-flux imposes the constraint
\begin{equation}
 R^{\mu,\nu\rho\alpha\beta}\, p_\mu=0 \ , \label{eq:RMMomentumConstraint}
\end{equation}
which as argued in \cite{Gunaydin:2016axc} can be understood by duality from the twisted torus, wherein a two-cycle becomes homologically trivial and thus no M2-branes can wrap this cycle. However, this would-be M2-brane wrapping mode is related by U-duality to the momentum singled out by \eqref{eq:RMMomentumConstraint} which thus has to vanish.

In the string theory limit, where we compactify one of the spatial directions on a circle and consider the limit where its radius vanishes, the constraint \eqref{eq:RMMomentumConstraint} implies that there can be no D0-branes in the locally non-geometric string background. This, in turn, is related by T-duality along the other three directions to the Freed-Witten anomaly: After the three dualities, the locally non-geometric background becomes a three-torus $T^3$ with $H$-flux while the D0-brane would be a D3-brane wrapping this $T^3$. However, as $T^3$ is a spin$^c$ manifold, the Freed-Witten anomaly forbids precisely this configuration and thus is a dual version of momentum constraint \eqref{eq:RMMomentumConstraint}. This also agrees with quantitative expectations from nonassociative quantum mechanics on the $R$-flux background~\cite{Mylonas:2013jha}.

\begin{figure}[t] 
\centering
  \includegraphics{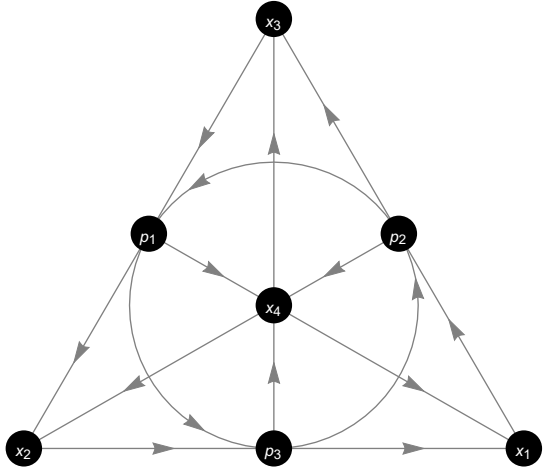}
  \caption{Commutation relations of the nonassociative $R$-flux algebra. The commutator of two phase space coordinates is given by the third node along the line connecting the two. If the arrow points from the first to the second element of the commutator, one has a factor of $\ii$ on the right-hand side, otherwise $-\ii$.}\label{f:MagicR}
\end{figure}

For the particular choice of $R$-flux wherein the only non-vanishing tensor components are $R^{4,\nu\rho\alpha\beta}=R\, \varepsilon^{\nu\rho\alpha\beta}$, the constraint \eqref{eq:RMMomentumConstraint} reads
\bea
R\, p_4=0 \ ,
\label{eq:Rp40}\eea
and the corresponding 3-algebra adapted to this frame is given by~\cite{Kupriyanov:2017oob}
\small
\begin{equation}\label{eq:M3algebra}
\begin{split}
[x^i,x^j,x^k] &=\, -\frac{\ell_s^2}{2}  \, R\, \varepsilon^{ijk}\, x^4 \,, \qquad [x^i,x^j,x^4] = \frac{\lambda^2\,\ell_s^2}{2} \, R\,\varepsilon^{ijk}\, x_k \,, \\[4pt]
[p^i,x^j,x^k] &=\, -\frac{\lambda\,\ell_s^2}{2} \, R\, \varepsilon^{ijk}\, p_4-\frac{\lambda\,\ell_s^2}{2} \, R\, \big(\delta^{ij}\, p^k - \delta^{ik}\, p^j\big) \,, \\[4pt]
[p_i,x^j,x^4] &=\, - \frac{\lambda^2\,\ell_s^2}{2} \, R\, \delta_i^j\, p_4-\frac{\lambda^2\,\ell_s^2}{2} \, R\, \varepsilon^{ijk}\, p_k \,,\\[4pt]
[p_i,p_j,x^k] &=\, \frac{\hbar^2\,\lambda}{2\,\ell_s^2} \, \varepsilon_{ij}{}^k\, x^4+\frac{\hbar^2\,\lambda^2}{2\, \ell_s^2}\,\big(\delta_j^k\, x_i-\delta_i^k\, x_j\big) \,, \\[4pt]
[p_i,p_j,x^4] &=\, -\frac{\hbar^2\,\lambda^3}{2\, \ell_s^2} \, \varepsilon_{ijk}\, x^k \,, \qquad
[p_i,p_j,p_k] = -\frac{2\,\hbar^2\,\lambda^2}{\ell_s^2} \, \varepsilon_{ijk}\, p_4 \,, \\[4pt]
[p_4,x^i,x^j] &=\, \frac{\lambda\, \ell_s^2}{2} \, R\, \varepsilon^{ijk}\, p_k \,, \qquad
[p_4,x^i,x^4] = -\frac{\lambda^2\, \ell_s^2}{2} \, R \, p^i \,, \\[4pt]
[p_4,p_i,x^j] &=\, -\frac{\hbar^2\,\lambda}{2\, \ell_s^2} \, \delta_i^j\, x^4 - \frac{\hbar^2\,\lambda^2}{2\, \ell_s^2}  \, \varepsilon_i{}^{jk}\, x_k \,, \\[4pt]
[p_4,p_i,x^4] &=\, -\frac{\hbar^2\,\lambda^3}{2\, \ell_s^2}  \, x_i \,, \qquad
[p_4,p_i,p_j] \ = \ -\frac{\hbar^2\,\lambda^2}{2\, \ell_s^2}  \, \varepsilon_{ijk}\, p^k \,,
\end{split}
\end{equation}
\normalsize
where the parameter $\lambda$ is related to the radius of the M-theory circle parameterised by $x^4$ and is specified more precisely below. This 3-algebra is understood to be supplemented by the constraint \eqref{eq:Rp40}. For $R\neq0$, this forces $p_4=0$, and since $p_4$ is a dynamical variable we need to implement this constraint following the gauge-fixing procedure described above. Thus we set $f(x,p)=-\frac2{\ii\hbar\,\lambda}\, p_4$, which breaks the symmetry group of the seven-dimensional phase space to $G_2 \supseteq SO(4) \simeq SU(2) \times SU(2)$, where $SO(4)$ is the subgroup leaving invariant the split into position and momentum coordinates. The corresponding nonassociative algebra is given by~\cite{Gunaydin:2016axc}
\begin{equation}
\begin{split}
[x^i,x^j] &=\, {\frac{\ii\ell_s^3}\hbar \,R \, \varepsilon^{ijk}\, p_k \,, \qquad
[x^4,x^i] =\, \frac{\ii \lambda \,\ell_s^3}\hbar\, R\, p^i \,, 
\qquad [p_i,x^4] =\, \frac{\ii\hbar\, \lambda^2}{\ell_s} \,x_i} \,, \\[4pt]
 \,, 
[x^i,p_j] &=\, {\frac{\ii\hbar}{\ell_s}\, \delta_{j}^i\, x^4+\frac{\ii\hbar\, \lambda}{\ell_s}\, \varepsilon^i{}_{jk}\, x^k \,, \qquad [p_i,p_j] =\, \frac{-\ii\hbar\,\lambda}{\ell_s}\, \varepsilon_{ijk}\, p^k} \,,
\end{split}
\label{eq:Ralgebra}
\end{equation}
whose Jacobiators are given by setting $p_4=0$ in the remaining 3-brackets from \eqref{eq:M3algebra}. This algebra, with $R = \lambda = \hbar = \ell_s = 1$, can be conveniently represented as in Figure~\ref{f:MagicR}, where each node is labelled by a phase space coordinate.

The 3-algebra provides a ``covariant'' formulation of the phase space algebra \eqref{eq:Ralgebra}, whereas imposing the constraint $R^{\mu;\nu\rho\alpha\beta}\, p_\mu=0$ explicitly, e.g. by setting $p_4=0$, obscures some of the symmetry. Understanding the action of the constraint \eqref{eq:RMMomentumConstraint} on the position and momentum coordinates therefore requires the symmetry breaking pattern $Spin(7) \longrightarrow SU(2)^3 \longrightarrow SO(4) \subseteq G_2$. Under the embedding $SU(2)^3 \subseteq Spin(7)$, the spinor representation $\bf 8$ restricts according to the decomposition
\bea
{\bf 8}\big|_{SU(2)^3} = \mbf{(2,1,2)} \ \oplus \ \mbf{(1,2,2)} \ ,
\eea
corresponding to the four position and four momentum coordinates. To implement the phase space constraint we break $SU(2)^3 \longrightarrow SO(4)$ by identifying two of the $SU(2)$ factors with their diagonal subgroup. One then obtains the representations
\begin{equation}
 {\mbf 8}\big|_{SO(4)} = \mbf{(2,2)} \ \oplus \ \mbf{(1,1)} \ \oplus \ \mbf{(1,3)} \ .
\end{equation}
Here we recognise the anticipated $SO(4)$ representation content of the reduced M2-brane phase space from~\cite{Gunaydin:2016axc}: $x^\mu$, $\mu=1,2,3,4$ transform in the fundamental representation $\mbf{(2,2)}$, $p_i$, $i=1,2,3$ in the chiral representation $\mbf{(1,3)}$, and the remaining momentum $p_4$ in the trivial representation $\mbf{(1,1)}$; in particular, the latter can be used to represent a central element, as in~\cite{Kupriyanov:2017oob}, and therefore to impose the constraint $p_4=0$.

In the contraction limit $\lambda=0$, which reduces M-theory to IIA string theory, the algebra (\ref{eq:Ralgebra}) reduces to the commutation relations for
the nonassociative phase space algebra describing the motion of a string
in an $R$-flux background:
\beq
\begin{split}
[x^i,x^j] &=\, {\frac{\ii\ell_s^3}\hbar  \,R \, \varepsilon^{ijk}\, p_k }
                  \ , \\[4pt]
{} [x^i,p_j] &=\, {\ii\hbar\, \delta^i_{j}}\ , \\[4pt]
[p_i,p_j] &=\, 0 \ ,
\end{split}
\label{eq:Ralgebrazero}\eeq
with the non-vanishing Jacobiators
\bea
[\![x^i,x^j,x^k]\!] = \ell_s^3\, R\, \varepsilon^{ijk} \ .
\eea
It is important to note that in the contraction limit $\lambda=0$, the M-theory direction $x^4$ becomes a central element of the contracted algebra, which can therefore be set to any constant value, that here we take to be $x^4=\ell_s$ since $x^4$ has units of length. As already mentioned in \S\ref{intro}, the nonassociativity of this phase space algebra is due to the locally non-geometric nature of the $R$-flux background.

\subsection{$Spin(7)$ automorphisms and magnetic dual background\label{sec:magneticdual}}

Now we wish to find the magnetic field analogue of the M-theory $R$-flux
algebra. We call this the \emph{magnetic dual algebra} and define it by interchanging the roles of position and
momentum coordinates in the 3-bracket on the eight-dimensional phase space. 
More precisely, this exchange is provided by a standard canonical
transformation of order four between position and momentum variables of the form
\begin{equation}
 \frac1{\ell_s} x_R^\mu \, \longmapsto \, - \frac{\ell_m}{\hbar} p_{M\,\mu} \ , \qquad \ell_s\, p_{R\,\mu} \, \longmapsto \, \frac{\hbar}{\ell_m} x_M^\mu \ ,
\label{eq:cantransf}\end{equation}
where again the subscripts $R$ and $M$ distinguish the $R$-flux and magnetic phase space variables, together with the replacement of the $R$-flux by a uniform magnetic charge density $\rho=\nabla\cdot\vec B = \frac{N}{\ell_m^3}$ through
\begin{equation}
 \frac{N}{\hbar} \longmapsto \, R \,.
\end{equation}
This canonical transformation preserves the M2-brane 3-algebra due to self-duality of the four-form $\phi_{\hat A\hat B\hat C\hat D}$ in eight dimensions~\cite{Kupriyanov:2017oob}:
\bea
\varepsilon_{\hat A\hat B\hat C\hat D\hat E\hat F\hat G\hat H}\, \phi_{\hat E\hat F\hat G\hat H} = \phi_{\hat A\hat B\hat C\hat D} \ ,
\eea
and it maps the 3-brackets \eqref{eq:M3algebra} to
\small
\begin{equation}\label{eq:dualM3algebra}
\begin{split}
[x^i,x^j,x^k] &=\, -2\, \lambda^2 \, \ell_m^2 \varepsilon^{ijk}\, x^4 \,, \qquad
[x^i,x^j,x^4] = - \frac{\lambda^2 \ell_m^2}{2} \, \varepsilon^{ijk}\, x_k \,, \\[4pt]
[p^i,x^j,x^k] &=\, \frac{\lambda\,\ell_m ^2}{2} \, \varepsilon^{ijk}\, p_4 + \frac{\lambda^2 \ell_m^2}{2} \, \big(\delta^{ij}\, p^k - \delta^{ik}\,p^j \big) \,, \\[4pt]
[p_i,x^j,x^4] &=\, - \frac{\ell_m^2\,\lambda}{2} \, \delta_i^j\, p_4-\frac{\ell_m^2\, \lambda^2}{2} \, \varepsilon^{ijk}\, p_k \,, \\[4pt]
[p_i,p_j,x^k] &=\, -\frac{\hbar\,\lambda}{2\,\ell_m^2} N \, \varepsilon_{ij}{}^k\, x^4 + \frac{\lambda\, N}{2 \ell_m^2} \big(\delta_j^k\, x_i-\delta_i^k\, x_j\big) \,, \\[4pt]
[p_i,p_j,x^4] &=\, \frac{\hbar\, \lambda\, N}{2} \varepsilon_{ijk}\, x^k \,, \qquad 
[p_i,p_j,p_k] = -\frac{\hbar\,N}{2\, \ell_m^2}\, N \, \varepsilon_{ijk}\, p_4 \,, \\[4pt]
[p_4,x^i,x^j] &=\, - \frac{\ell_m^2\, \lambda^3}{2} \, \varepsilon^{ijk}\, p_k \,, \qquad
[p_4,x^i,x^4] = \frac{\ell_m^2\,\lambda^3}{2} \, p^i \,, \\[4pt]
[p_4,p_i,x^j] &=\, \frac{\hbar\,\lambda^2\,N}{2\,\ell_m^2} \, \delta_i^j\, x^4 - \frac{\hbar\,\lambda^2\,N}{2\,\ell_m^2} \, \varepsilon_i{}^{jk}\, x_k \,, \\[4pt]
[p_4,p_i,x^4] &=\, \frac{\hbar\,\lambda^2\,N}{2} \, x_i \,, \qquad
[p_4,p_i,p_j] = \frac{\hbar\,\lambda^2}{2\,\ell_m^2} \, N\, \varepsilon_{ijk}\, p^k \,.
\end{split}
\end{equation}
\normalsize
In turn, this corresponds to an automorphism of $Spin(7)$ which exchanges two of the $SU(2)$ factors in the embedding $SU(2)^3\subseteq Spin(7)$. Now $p_\mu$ transform in the fundamental representation of $SO(4)$, $x^i$ in the chiral representation, and $x^4$ in the trivial representation; in particular, we can now impose the constraint $x^4=0$ to reduce to a seven-dimensional phase space with quasi-Poisson structure which is ``dual'' to that of the reduced M2-brane phase space.

\begin{figure}[t]
\centering
  \includegraphics{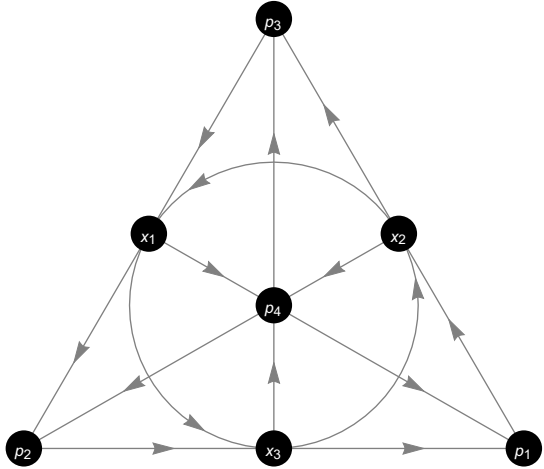}
  \caption{Commutation relations of the nonassociative magnetic monopole algebra. The commutator of two phase space coordinates is given by the third node along the line connecting the two. If the arrow points from the first to the second element of the commutator, one has a factor of $\ii$ on the right-hand side, otherwise $-\ii$.}\label{f:MagicM}
\vspace{-10pt}
\end{figure}

Under this canonical transformation, the $R$-flux algebra (\ref{eq:Ralgebra}) turns into the magnetic dual phase space algebra given by
\begin{equation}
\begin{split}
[x^i,x^j] &=\, -\ii\ell_m\,\lambda\, \varepsilon^{ijk}\, x_k \,, \\[4pt]
[p_i,x^j] &=\, \ii \ell_m\, \delta_{i}^j\, p_4+\ii\ell_m\, \lambda\, \varepsilon_i{}^{jk}\, p_k \,, \qquad
[x^i,p_4] = \ii \ell_m\, \lambda^2\, p^i \,, \\[4pt]
[p_i,p_j] &=\, \frac{\ii\hbar\,N}{\ell_m^3} \, \varepsilon_{ijk}\, x^k \,, \qquad
[p_4,p_i] = \frac{\ii\hbar\,\lambda\,N}{\ell_m^3} \, x_i \,,
\end{split}
\label{eq:magneticalgebra}
\end{equation}
which also follows from the nonassociative octonion algebra, and can again be represented diagrammatically as in Figure~\ref{f:MagicM}. The canonical transformation has a natural realisation on the diagram: it acts  by exchanging vertices with the opposite faces, and replacing $x_4 \longleftrightarrow p_4$.

When
$\lambda=0$ these are just the commutation relations for the nonassociative algebra describing the motion of a charged particle
in a uniform magnetic monopole background:
\beq
\begin{split}
[x^i,x^j] &=\, 0 \,, \\[4pt]
[x^i,p_j] &=\, \ii\hbar\, \delta^i_{j} \,, \\[4pt]
[p_i,p_j] &=\, \frac{\ii\hbar\,N}{\ell_m^3}  \, \varepsilon_{ijk}\, x^k \,,
\end{split}\label{eq:magneticalgebrazero}\eeq
with the non-vanishing Jacobiators
\bea
[\![p_i,p_j,p_k]\!] = -\frac{\hbar^2\,N}{\ell_m^3}\,\varepsilon_{ijk} \,.
\eea
Again in the contraction limit $\lambda=0$, the momentum $p_{4}$ becomes a central element of the contracted algebra, which can therefore be set to any constant value, that here we take to be $p_{4}=-\frac{\hbar}{\ell_m}$ for dimensional reasons.

\vskip1em

\subsection{Monopoles and quantum gravity\label{sec:mongravity}}

We can gain some insight into the meaning of the position space noncommutativity for $\lambda\neq0$ by observing that for $N=0$ the algebra \eqref{eq:magneticalgebra} reveals a noncommutative but associative deformation of position space with $[p_\mu,p_\nu]=0$. We can show that
\begin{equation}
 r^2 := \lambda^2 \, p_i \, p^i + p_4^2 
\end{equation}
is a central element of this algebra; this follows from $\left[ r^2, p_i \right] = \left[r^2, p_4 \right] = 0$ and 
\begin{equation}
  \left[ p_4^2, x^i \right] = - \lambda^2 \, \left[ p_j\, p^j , x^i
  \right] = - \ii \hbar \, \lambda^2\, \left( p^i \, p_4 + p_4\, p^i
  \right) \ .
\end{equation}
We can therefore restrict the momenta to a unit three-sphere $S^3\subseteq \IR^4$ by taking $q_i = p_i / r$ and $q_4 = p_4/r$ so that
$(q_\mu)\in S^3$, i.e. $\lambda^2\, {\vec q}\,^2+q_4^2=1$. Under this
restriction the remaining commutation relations can be expressed on
the lower hemisphere $q_4\leq0$, $|\vec q\,|\leq\frac1\lambda$ of a
three-dimensional momentum space $S^3$ as
\begin{equation}
\begin{split}
[x^i,x^j] &=\, -\ii\ell_m\, \lambda\, \varepsilon^{ijk}\,
x_k\ , \\[4pt]  
[x^i,q_j] &=\,
\ii\ell_m\,
\sqrt{1-\lambda^2\, |\vec q\,|^2}\ \delta_{j}^i +\ii\ell_m\,
\lambda\, \varepsilon^i{}_{jk}\, q^k \ , \\[4pt]
[q_i,q_j]&=\, 0 \ .
\end{split}
\label{eq:QGalgebra}
\end{equation}

These are precisely the commutation relations that arise in a Ponzano-Regge spin foam model of three-dimensional quantum gravity~\cite{Freidel:2005me}: Integrating out the gravitational degrees of freedom there coupled to spinless matter yields an effective scalar field theory on an $SU(2)$ Lie algebra noncommutative position space. The momentum space in this case is curved and the position coordinates can be interpreted as right-invariant derivations on $SU(2)\simeq S^3$, so that noncommutativity in this instance is a consequence of the fact that massive particles coupled to three-dimensional gravity have momenta bounded by $\frac1\lambda$. The additional deformation of the canonical position-momentum commutator makes the algebra \eqref{eq:QGalgebra} symmetric under a $\kappa$-deformation of the Poincar\'e group.

This suggests that the magnetic dual of the M-theory $R$-flux algebra admits an interpretation in terms of monopoles in the spacetime of Euclidean quantum gravity in three dimensions, with the contraction parameter $\lambda$ identified with the Planck length $\ell_{\rm P}$ through 
\bea
\lambda=\frac{\ell_{\rm P}}{\ell_m} \,,
\label{eq:lambdaellP}\eea
where $\ell_m$ plays the role of Planck's constant. In this context the contraction limit $\lambda\to0$ corresponds to neglecting quantum gravitational effects, which freezes the extra momentum mode $q_4=\pm\,1$ and decompactifies the momentum space $\vec q\in\IR^3$ to get a canonical six-dimensional phase space. As $p_{4}$ varies, the phase space algebra \eqref{eq:magneticalgebra} at $N=0$ describes a cone over the noncommutative space defined by the algebra \eqref{eq:QGalgebra}, foliated by momentum space spheres $S^3$ of radius $r$.

Of course, we are not literally allowed to take the limit $N=0$ in the reduced seven-dimensional phase space, as it is the presence of magnetic charge that forces the constraint $x^4=0$ in the first place. In the following we will propose a precise M-theory background which realises this picture, and moreover describes the duality between the nonassociative $R$-flux algebra (\ref{eq:Ralgebra}) and the magnetic monopole algebra (\ref{eq:magneticalgebra}), which both follow identically from the nonassociative algebra of the octonions. We will show that in both dual algebras the meaning of the contraction parameter is the same, being identified as the Planck length through \eqref{eq:lambdaellP}.

\section{M-wave phase space}

\subsection{Non-geometric Kaluza-Klein monopole in M-theory\label{sec:KKmon}}

In this section we shall propose that the nonassociative magnetic dual
background discussed in \S\ref{sec:magneticdual} is given by a locally
non-geometric variant of the Kaluza-Klein monopole solution in M-theory~\cite{Townsend:1995kk}. It is defined by
the metric
\bea\label{metricTN}
\dd s^2 = \dd s_{M_7}^2 +R_{11} U\, \dd\vec y\cdot\dd\vec y + R_{11}\, U^{-1}\,\big( \dd z+\vec A\cdot\dd \vec y\, \big)^2 \ ,
\eea
where $\dd s_{M_7}^2$ is the metric of a seven-manifold $M_7$ and $(\vec y,z)\in \IR^3\times S^1$ are, generally, local coordinates on a hyper-K\"ahler four-manifold with a $U(1)$ isometry that preserves the hyper-K\"ahler structure; here $R_{11}$ is the asymptotic radius of the fibre $S^1$, $U(\vec y\,)$ is a harmonic function on $\IR^3$, and $A=\vec A(\vec y\,)\cdot\dd\vec y$ is a $U(1)$ gauge connection on $\IR^3$ whose curvature $F=\dd A$ is related by
\begin{equation}
 F = \ast_3\, \dd U \ ,
\end{equation}
where $\ast_3$ is the restriction of the 11-dimensional Hodge duality operator to the space $\mathbb{R}^3$ parameterised by $\vec{y}$. The hyper-K\"ahler structure will ensure that the metric \eqref{metricTN} yields half-BPS solutions.

Let us begin by reviewing the standard Kaluza-Klein
monopole solution. It is based on the seven-manifold $M_7=\IR^{1,6}$, and the usual point-like Dirac
monopole with magnetic charge $N\in\IZ_{>0}$ that is associated to a
minimal locally well-defined gauge connection $ A^{\rm D}$ and corresponding curvature $F^{\rm D}$ of the form
\begin{equation}
A_i^{\rm D}(\vec y\, )=\varepsilon_{ijk}\, {y^j\over |\vec y\,|}
\, \frac N{
  y_k+|\vec y\,| } \ , \qquad F^{\rm D}_{ij}(\vec y\,)=N\, \varepsilon_{ijk}\, {y^k\over |\vec y\,|^3}\ .
\end{equation}
This gauge field is singular along the negative coordinate axes for $y^k\leq0$; we can shift this coordinate singularity to the positive coordinate axes for $y^k\geq0$ by the gauge transformation $\chi(\vec y\,) =\varepsilon_{ijk}\, y^k\, \tan^{-1}\big(\frac{y^j}{y^i}\big)$ to get the connection
\bea
A^{\rm D}+\dd\chi = \varepsilon_{ijk}\, \frac{y^j}{|\vec y\,|}\,
\frac{N\, \dd y^i}{y_k-|\vec y\,|} \ .
\eea
It indeed leads to a point-like magnetic monopole source with charge density
\begin{equation}
\rho^{\rm D}(\vec y\,)= \nabla \cdot \vec{B}(\vec y\,) = 4\pi\, N\, \delta^{(3)}(\vec y\, ) \ ,
\end{equation}
and the Dirac monopole can be described as a connection on a
$U(1)$-bundle of first Chern class $N$ over $\IR^3\setminus\{\vec0\,\}$, which is homotopic to $S^2$.
The asymptocially (locally) flat harmonic function is
\bea
U(\vec y\,) = \frac {N}{|\vec y\,|} + \frac{1}{R_{11}} \ ,
\eea
and it realises the Euclidean Taub-NUT space as a local $S^1$-fibration over $\IR^3$ of degree $N$. More precisely, in the region outside the degeneration locus at the origin, where the monopole is situated and the $S^1$ fibre shrinks to the fixed point of an ${\sf A}_{N-1}$ orbifold singularity, the space $\IR^3\setminus\{\vec0\,\}$ is homotopic to $S^2$ and the Taub-NUT circle bundle is homotopic to the Seifert fibration over the embedding $S^2\hookrightarrow\IR^3\setminus\{\vec0\,\}$ with the lens space $S^3/\IZ_N$ as total space; for $N=1$ this is just the familiar Hopf fibration $S^3\to S^2$. 

In the present case of interest we require a constant magnetic charge distribution $\rho$:
\begin{equation}
\rho= \nabla \cdot \vec{B} = \ast_3\, \dd F = \Box_3\, U = \frac{N}{\ell_m^3} \ ,
\end{equation}
where again $N\in\IZ_{>0}$ by (generalised) Dirac charge quantization of the
corresponding dual fluxes in string theory.
This is solved by
\begin{equation}
 U(\vec y\,) = \frac{N}{2\,\ell_m^3} \, |\vec y\,|^2 + \frac1{R_{11}} \,,
\end{equation}
which is of course not asymptotically (locally) flat since we consider a uniform magnetic charge distribution $\rho$, but it is flat near the origin of $\IR^3$. The appropriate magnetic field, which is linear in $\vec y$, is also easily determined:

\begin{equation}
F_{ij}(\vec y\,)=\varepsilon_{ijk}\, B^k(\vec y\,) ={N\over 3\,\ell_m^3} \, \varepsilon_{ijk}\, y^k \ .
\end{equation}
This is analogous to the fact that the dual non-geometric $R$-fluxes are generated by linear multi-vectors. Now since the two-form $F$ is nowhere closed, the associated gauge field $\vec A$ does not even exist locally. Hence there is also no local expression for the metric (\ref{metricTN}). This is due to the fact that the constant magnetic charge distribution is globally defined over all of $\IR^3$, so that now there is no need to excise the origin. The solution describes a $U(1)$-gerbe over $\IR^3$, rather than a $U(1)$-bundle over $\IR^3\setminus\{\vec0\,\}$, and since $\IR^3$ is contractible the gerbe is necessarily trivial. However, the trivial gerbe can still have a non-zero curving which is given by the two-form $F$ whose curvature is $H= \dd F=\ast_3\, \rho$. While $H$ is non-zero, its cohomology class vanishes because the gerbe is trivial. Nevertheless, the total space of the trivial gerbe on $\IR^3$ is simply $\IR^3\times S^1$, which has exactly the local geometry needed to insert a Kaluza-Klein type monopole solution into the four-dimensional part of the metric \eqref{metricTN}.

What is then the meaning of the metric \eqref{metricTN} in this case?
Let us first give a heuristic argument, which produces at least a
formal non-local smeared expression for $\vec A$ in the following way.
We know that the uniform magnetic charge distribution can be obtained via an integral over infinitely-many densely distributed Dirac monopoles:
\begin{equation}
\rho={1\over 4\pi \ell_m^3}\, \int\, \rho^{\rm D}(\vec y-\vec y\,') \ \dd^3\vec y\,'=\frac{N}{\ell_m^3}\ .
\label{eq:rhointN}\end{equation}
The same is true for the magnetic field:
\begin{equation}
F_{ij}(\vec y\,)={1\over 4\pi\ell_m^3}\, \int\, F_{ij}^{\rm D}(\vec y-\vec y\,') \ \dd^3\vec y\,'={N\over 3\ell_m^3}\, \varepsilon_{ijk}\, y^k\ .
\label{eq:Fint}\end{equation}
The integration here can be defined more precisely by using
distributions supported on a
ball centred at $\vec y$ in the large radius limit, or alternatively
by formally applying Stokes' theorem.
Therefore we can express the gauge field in the non-local smeared form
\bea
A_{i}(\vec y\, ) = {1\over 4\pi\ell_m^3}\, \int\, A_{i}^{\rm D}(\vec y-\vec
y\,') \ \dd^3\vec y\,' = {N\over 4\pi\ell_m^3} \,\varepsilon_{ijk}\, \int\, 
{(y_j-y_j')\over |\vec y-\vec y\,'|\, \big((y_k-y_k')+|\vec y-\vec
  y\,'|\big)} \ \dd^3\vec y\,' \ ,
\label{eq:nonlocalA}\eea
where here the integration should be additionally defined so that it avoids
the Dirac string singularities. The fact that for the case of constant
magnetic charge density there is no local expression for the gauge field $\vec A$, and hence no local expression for the
Kaluza-Klein monopole metric, is analogous to the observation that for the $R$-flux
background there is also no local expression for the metric.

It is instructive to approximate the constant
magnetic charge distribution by a finite sum of Dirac monopoles and hence to compare with
the standard Kaluza-Klein monopole solution based on the
multi-centred version of the Taub-NUT geometry~\cite{Hawking:1976jb,Gibbons:1996nt}, which is
defined by the asymptotically (locally) flat harmonic function
\bea
U(\vec y\,) = \sum_{a=1}^K \, \frac {N}{|\vec y-\vec y_a|} + \frac1{R_{11}} \ .
\label{eq:multicentredTN}\eea
It corresponds to a magnetic charge distribution consisting of $K$
Dirac monopoles situated at points $\vec
y_a\in\IR^3$, with gauge field
\bea
\vec A(\vec y\,) = \sum_{a=1}^K \, \vec
A\,^{\rm D}(\vec y-\vec y_a) \ . 
\eea
The multi-centred Taub-NUT space is again locally the same
$S^1$-fibration over $\IR^3$, where now the $S^1$ fiber shrinks to an
${\sf A}_{N-1}$ orbifold fixed point over each monopole location $\vec y_a$, $a=1,\dots,K$. When the position moduli $\vec y_a$ are generic, the
space has $K-1$ homologically non-trivial two-cycles on which M2-branes can
wrap; they are constructed as the inverse images, under the $S^1$-bundle projection, of paths connecting neighbouring singular points $\vec y_{a}, \vec y_{a+1}$, $a=1,\dots,K-1$ in $\IR^3$. When two or more of the $\vec y_a$ coincide, the corresponding two-cycles collapse and additional ${\sf A}$-type orbifold singularities develop. Hence the smeared solution defined above, which uses a continuous distribution of Dirac monopoles, becomes highly singular because any pair of monopoles can come arbitrarily close to one another; in particular, there are no M2-brane wrapping modes in this background.

To understand more precisely the geometric meaning of the smeared Kaluza-Klein monopole, we need a geometric interpretation of the Dirac monopole over all of $\IR^3$ rather than simply on the excised space $\IR^3\setminus\{\vec0\,\}$. For this,
we appeal to a construction
from~\cite{Brylinski:1993ab} (see also~\cite{Pande:2006id}). The
crucial observation is that the magnetic charge of a Dirac monopole
can be interpreted as the cohomology class of the closed three-form
delta-function $H^{\rm D}= \ast_3\,\rho^{\rm D}$ supported at $\vec0$ by virtue of
Maxwell's equations with magnetic sources: $\dd F^{\rm
  D}=\ast_3\,\rho^{\rm D}=H^{\rm D}$; this class trivially vanishes as
an element of $H^3(\IR^3;\IZ)=0$. However, since the two-form $F^{\rm
  D}$ cannot be extended through the origin of $\IR^3$, the class is
non-trivial as an element of the relative cohomology
$H^3(\IR^3,\IR^3\setminus\{\vec0\,\};\IZ)$ generated by three-forms on
$\IR^3$ with singular support at $\vec 0\,$; the expression
\eqref{eq:rhointN} with $\vec y=\vec0$ then means that the cohomology class of $H^{\rm D}$ in $
H^3(\IR^3,\IR^3\setminus\{\vec0\,\};\IZ)=\IZ$ is the magnetic charge $N$. To see that this description of the monopole charge is equivalent to that above as the first Chern class of a $U(1)$-bundle over $\IR^3\setminus\{\vec 0\,\}$, which is a class in $H^2(\IR^3\setminus\{\vec 0\,\};\IZ)=\IZ$, we use the long exact cohomology
sequence for the pair $(\IR^3,\IR^3\setminus\{\vec0\,\})$ and the fact
that all cohomology groups of $\IR^3$ vanish to show that there is an isomorphism
\bea
H^2\big(\IR^3\setminus\{\vec0\,\};\IZ\big)\xrightarrow{ \ \simeq \ }
H^3\big(\IR^3,\IR^3\setminus\{\vec0\,\};\IZ \big)
\eea
which sends the class of $F^{\rm D}$ to the class of $\dd F^{\rm D}=H^{\rm D}$. 

In this interpretation we can
move the Dirac monopole around in $\IR^3$, or inside its one-point
compactification which is the three-sphere $S^3=\IR^3\cup\{\infty\}$. For this, we observe that the embedding
$\IR^3\hookrightarrow S^3$ induces an isomorphism~\cite{Pande:2006id}
\bea
H^3\big(\IR^3,\IR^3\setminus\{\vec0\,\};\IZ \big) \simeq
H^3\big(S^3,S^3\setminus\{\vec0\,\};\IZ \big) \xrightarrow{\ \simeq \
} H^3\big(S^3;\IZ \big)=\IZ \ ,
\eea
where the first isomorphism follows from the excision theorem and the second from the fact that $S^3\setminus\{\vec0\,\}$ is contractible. In other
words, the usual geometric
description of the Dirac monopole as a $U(1)$-bundle with connection,
whose curvature two-form represents a class in
$H^2(\IR^3\setminus\{\vec0\,\};\IZ)$, can be equivalently described by
a $U(1)$-gerbe with connective structure whose curvature three-form
represents a class in $H^3(S^3;\IZ)$; this means that we now deal directly with the nonassociativity induced by the point-like magnetic charge, which is exactly what we need for our purposes. In particular, the same class is
obtained if we move the location of the monopole from the origin to
any other point $\vec y\,'\in\IR^3$.

The information about the location $\vec y\,'$ of the monopole is entirely encoded in the connective structure on this gerbe, which can be described explicitly in terms of the monopole connection $A^{\rm D}_{\vec y\,'}=\vec A\,^{\rm D}(\vec y-\vec y\,')\cdot \dd\vec y$ on $\IR^3\setminus\{\vec y\,'\}$. For this, we take a stereographic cover of $S^3$ with contractible open sets
$U_{\vec y\,'}= S^3\setminus\{\vec y\,'\}$ and
$U_\infty=S^3\setminus\{\infty\} = \IR^3$, whose intersection is
$U_{\vec y\,',\infty}:= U_{\vec y\,'}\cap U_\infty=S^3\setminus\{\vec y\,',\infty\}= \IR^3\setminus\{\vec y\,'\}$. Choose a
partition of unity $\chi_{\vec y\,'},\chi_\infty$ subordinate to the
cover $U_{\vec y\,'},U_\infty$, i.e. $\chi_{\vec y\,'},\chi_\infty$ are
functions on $S^3$ supported in $U_{\vec y\,'},U_\infty$ which satisfy
\bea
\chi_{\vec
  y\,'}+\chi_\infty=1 \ .
\eea
Define curving two-forms $b_{\vec y\,'}=-\chi_\infty\,
F_{\vec y\,'}^{\rm D}$ on $S^3\setminus\{\vec y\,'\}$ and $b_\infty=\chi_{\vec y\,'} \, F_{\vec y\,'}^{\rm D}$ on
$S^3\setminus\{\infty\}=\IR^3$, so that $b_{\infty}-b_{\vec y\,'}=F_{\vec y\,'}^{\rm D}=\dd A^{\rm D}_{\vec y\,'}$ on $U_{\vec
  y\,',\infty}=\IR^3\setminus\{\vec y\,'\}$. Since $F^{\rm D}_{\vec y\,'}$ is a closed two-form on $\IR^3\setminus\{\vec y\,'\}$, the curvature
three-form
\bea
h_{\vec y\,'}=\dd b_{\vec y\,'}=-\dd \chi_\infty\wedge F_{\vec y\,'}^{\rm
  D}=\dd \chi_{\vec y\,'}\wedge F_{\vec y\,'}^{\rm D}=\dd b_\infty
\eea
is globally defined
on $S^3$ but supported in $\IR^3\setminus\{\vec y\,'\}$. Transforming
to spherical coordinates on $\IR^3$ and choosing $\chi_{\vec y\,'}$ so
that it depends only on the square $r^2$ of the radial coordinate, by
using Stokes' theorem one computes~\cite{Brylinski:1993ab}
\bea
\int_{S^3}\, h_{\vec y\,'} = \int_{S_{\vec y\,'}^2}\, F^{\rm D}_{\vec
  y\,'} = 4\pi\,N \ ,
\eea
where $S_{\vec y\,'}^2$ is a sphere centred at $\vec y\,'\in\IR^3$; it follows that the Dixmier-Douady class of this gerbe in $H^3(S^3;\IZ)=\IZ$ is just the magnetic charge~$N$.

This construction suggests a geometric meaning for the expressions
\eqref{eq:rhointN}--\eqref{eq:nonlocalA}: They
describe the connective structure on the constant magnetic charge
gerbe on $\IR^3$ realised as a family of Dirac monopole gerbes,
parameterised by the monopole locations $\vec y\,'\in\IR^3$. In
particular, the inherent non-locality of the gauge field
\eqref{eq:nonlocalA} is reflected in the feature that only at each point $\vec y\,'\in\IR^3$ can we identify the connection $A^{\rm
  D}_{\vec y\,'}$ of the gauge
bundle of a Dirac monopole located at $\vec y\,'$, which is defined on $U_{\vec y\,',\infty}=\IR^3\setminus\{\vec y\,'\}$. This interpretation is
similar in spirit to the global description of non-geometric string
backgrounds as fibrations of
noncommutative and nonassociative tori~\cite{Bouwknegt:2004ap}, and
indeed we shall suggest below a more precise interpretation of this singular
Kaluza-Klein monopole solution in terms of nonassociative geometry
adapted to the family of Dirac monopole gerbes over~$\IR^3$.

\subsection{M-waves} \label{s:MwavePSpace}

One now needs to probe the background of \S\ref{sec:KKmon} by degrees of freedom that are
electrically charged under the $U(1)$ gauge field $\vec A$, which appears as a metric component in the 11-dimensional solution. That this
should be the M-wave (or pp-wave) along the $z$-direction can be seen
as follows. By reducing on the circle $S^1$ parameterised by $z$, the usual
Kaluza-Klein monopole of M-theory becomes a
D6-brane of IIA string
theory wrapped on $M_7$, which is magnetically dual to a
D0-brane and couples magnetically to $\vec A$. From the perspective of the
11-dimensional theory, the electric dual to the NUT-charge is thus a
graviton momentum mode. 

For our non-geometric Kaluza-Klein monopole, the absence of a local
gauge field $\vec A$ is not an issue on reduction to IIA string
theory, because only the magnetic field $F$ carries physically
relevant degrees of freedom, and the violation of the Bianchi identity
$\dd F=\ast_3\,\rho$ is now accounted for by a uniform  distribution
of smeared D6-brane sources. The meaning of the uplift of such
continuous distributions of D6-branes to M-theory is generally an open
problem which has been investigated only for some special instances
at the level of supergravity equations of motion, see
e.g.~\cite{Gaillard:2009kz,Danielsson:2014ria}. In these settings a
dual ``smeared'' Kaluza-Klein monopole in M-theory is interpreted as a
deformation of target space geometry by torsion, i.e. geometric
flux, which leads to source-modified equations of motion that match under M-theory/IIA duality; for instance, in certain cases the uplift involves taking $M_7$ to be a $G_2$-manifold in \eqref{metricTN}. Our approach to the current problem at hand is different, in that we interpret the smeared Kaluza-Klein monopole as sourcing a nonassociative 
deformation of
phase space geometry that we propose to be equivalent to the non-local geometric description of
\S\ref{sec:KKmon}.
All of this suggests that one should identify $x^i=y^i$ and $x^4=z$, such that our proposed phase space algebra of the non-geometric Kaluza-Klein monopole background in M-theory, which follows from the nonassociative octonion algebra, is given by \eqref{eq:magneticalgebra}. Let us now provide some evidence for this proposal.

As the appropriate probe of this background is a momentum wave along the $x^4$-direction, it has no local position with respect to $x^4$, and therefore the configuration space of the M-wave is described by the three coordinates $(x^1,x^2,x^3)$. However, the wave can have momenta in all four directions, and hence the phase space of the M-wave is seven-dimensional with coordinates $(x^1,x^2,x^3; p_1,p_2,p_3,p_4)$. This agrees with the seven-dimensional phase space that appears in the nonassociative algebra (\ref{eq:magneticalgebra}). The reduction from M-theory to string theory is described by the limit $g_s,R_{11}\rightarrow 0$ of weak string coupling and vanishing radius of the eleventh dimension. Hence we would like to identify the limit where the contraction   parameter $\lambda\rightarrow 0$ with the limit $g_s,R_{11}\rightarrow 0$.   The standard identification of M-theory parameters, which are the 11-dimensional Planck length $\ell_{\rm P}$ and the radius $R_{11}$, with the string theory parameters $\ell_s$ and $g_s$ is given as
\begin{equation}
 \ell_s^2={\ell_{\rm P}^3\over R_{11}} \ , \qquad
 g_s=\left({R_{11}\over \ell_{\rm P}} \right)^{3/2} \ .
\end{equation}
It follows that, in order to keep the string length $\ell_s$ finite, the limit  $R_{11}\rightarrow 0$ should be taken with the Planck length vanishing as $\ell_{\rm P}\sim R_{11}^{1/3}$, which implies that the string coupling vanishes as $g_s\sim R_{11}$. This corresponds to the contraction limit $\lambda \rightarrow 0$ if we make the identification proposed in \eqref{eq:lambdaellP}.
  
Let us also see what happens to the phase space variables of the
M-wave probe in this limit. In fact, the M-wave becomes an
electrically charged D0-particle when $R_{11}\rightarrow 0$. Its
momentum $p_4$ in the M-theory direction disappears from the phase
space and we end up with the six-dimensional phase space of the
electric particle in the magnetic monopole background, correctly described
by the coordinates $(x^1,x^2,x^3;
p_1,p_2,p_3)$. The disappearance of $p_4$ can be easily seen by noting that 
\begin{equation}
p_4={\hbar\, e\over R_{11}}\ ,
\label{eq:p4eR11}\end{equation}
so that for $R_{11}\rightarrow 0$ the momentum $p_4$ is frozen and the Kaluza-Klein momentum quantum number $e$  becomes the electric charge of the D0-particle.
 
\subsection{Free M-theory phase space} \label{s:8dPSpace}

As observed in~\cite{Kupriyanov:2017oob}, in the absence of non-geometric
flux, $R=0$, there is no longer any constraint imposed on the momentum
$p_4$ by \eqref{eq:Rp40},
and since the $R$-flux simply appears as a parameter one can set it to
zero in the 3-algebra. In this limit the brackets \eqref{eq:M3algebra} describe the free phase space 3-algebra structure of M2-branes with the non-vanishing 3-brackets
\beq \label{eq:Mphase3alg}
\begin{split}
[p_i,p_j,x^k] &=\, \frac{\hbar^2\,\lambda}{2\,\ell_s^2} \, \varepsilon_{ij}{}^k\, x^4+\frac{\hbar^2\,\lambda^2}{2\,\ell_s^2} \,\big(\delta_j^k\,x_i-\delta_i^k\, x_j\big) \,, \qquad
[p_i,p_j,x^4]  = -\frac{\hbar^2\,\lambda^3}{2\,\ell_s^2} \, \varepsilon_{ijk}\, x^k \,, \\[4pt] 
[p_i,p_j,p_k] &=\, \frac{-2\, \hbar^2\,\lambda^2}{\ell_s^2} \, \varepsilon_{ijk}\, p_4 \,, \qquad
[p_4,p_i,x^j] = -\frac{\hbar^2\,\lambda}{2\,\ell_s^2} \, \delta_i^j\, x^4- \frac{\hbar^2\,\lambda^2}{2\,\ell_s^2} \, \varepsilon_i{}^{jk}\, x_k \,, \\[4pt] 
[p_4,p_i,x^4] &=\, -\frac{\hbar^2\,\lambda^2}{2\,\ell_s^2} \, x_i \,, \qquad 
[p_4,p_i,p_j] \ = \ -\frac{\hbar^2\,\lambda^2}{2\,\ell_s^2} \, \varepsilon_{ijk}\, p^k \,.
\end{split}
\eeq
Here we expect the dimensionless quantity $\lambda = \lambda\left( \frac{\ell_{\rm P}}{\ell_s} \right)$ to vanish $\lambda \to 0$ as $\ell_{\rm P}\to 0$. Then all 3-brackets vanish in the limit $\ell_{\rm P}\to0$ where quantum gravitational effects are turned off. This algebra is symmetric under the $\IZ_4$-automorphism \eqref{eq:cantransf}, so that the same is true of the 3-algebra describing M-waves in the absence of magnetic charge, $\rho=\frac{N}{\ell_m^3}=0$ (with the roles of position and momentum coordinates interchanged in \eqref{eq:Mphase3alg}); in this limit the four-dimensional metric in the Kaluza-Klein monopole solution \eqref{metricTN} collapses to that of flat space $\IR^3\times S^1$. Thus in either limit $R\to0$ or $\rho\to0$ with finite $\ell_{\rm P}$, both eight-dimensional $Spin(7)$-symmetric 3-algebras are isomorphic and describe the unique nonassociative eight-dimensional phase space of M-theory.

Let us now consider the decompactification limit $R_{11}\to\infty$ of
the M-theory circle, so that now $x^4\in\IR$ and also
$p_4\in\IR$. From the M-theory perspective we can describe this more
covariantly as the limit of infinite volume $V \to \infty$. This is
the strong coupling limit of IIA string theory, wherein the
contraction parameter $\lambda$ (or equivalently the Planck length
$\ell_{\rm P}$) should nevertheless remain finite; this means that, as
a function of the string coupling $g_s$, the parameter $\lambda(g_s)$
should satisfy $\lambda(0)=0$ and $
\lambda(g_s)\to1$ as $g_s\to\infty$. Recall that one way to argue $R\neq0$ forces $p_4=0$
is based on the Freed-Witten anomaly, which in the present context
forbade the existence of D0-branes in the string $R$-flux background~\cite{Gunaydin:2016axc}; this implied $p_4=0$ by \eqref{eq:p4eR11}. However, in the decompactification limit the Freed-Witten anomaly argument no longer applies as the D0-particle number is only derived from M-theory as an electric Kaluza-Klein charge upon compactification, and the momentum $p_4\in\IR$ becomes a dynamical variable again.

In~\cite{Gunaydin:2016axc} the homology of the twisted torus was also
used to argue for the constraint \eqref{eq:Rp40}. However, from the
duality with the twisted torus, one necessarily obtains an $R$-flux
which scales as $V^{-1/2}$, with $V$ the volume of the four-torus
$T^4$ with $R$-flux. Thus we find that the configuration of
non-vanishing $R$-flux on a non-compact space is not related by U-duality to the
twisted torus. This means that the argument used to set $p_4 = 0$
again fails to hold in the decompactification limit $V \to
\infty$. Hence we can postulate that the constraint \eqref{eq:RMMomentumConstraint}
may in fact be more generally given by
\begin{equation}
 \frac1{V} \, R^{\mu,\nu\rho\alpha\beta}\, p_\mu = 0 \ ,
\end{equation}
which in the case considered in the present paper, where the only
non-zero component of the M-theory $R$-flux is $R^{4,\nu\rho\alpha\beta} = R \, \varepsilon^{\nu\rho\alpha\beta}$, becomes
\begin{equation}
 \frac{R \, p_4}{V} = 0 \ .
\end{equation}
In the decompactification limit $V \to \infty$, this allows for both
$R\neq0$ and $p_4\neq0$, and suggests a role of the M-theory 3-algebra
\eqref{eq:M3algebra} with both $p_4$ and $R$ non-vanishing.

Dually, in the non-geometric Kaluza-Klein monopole background, the decompactification limit of the direction along $x^4=z$ should now reinstate the local coordinate $x^4$ in the configuration space of the M-wave. For the standard multi-centred Taub-NUT space described by \eqref{eq:multicentredTN}, which near infinity looks like a circle bundle over $\IR^3$ with fibre $S^1$ of radius $R_{11}$, in the limit $R_{11}\to\infty$ it becomes an ALE space of type ${\sf A}_{K-1}$ that is asymptotic at infinity to $\IR^4/\IZ_K$, together with an ${\sf A}_{N-1}$ singularity at each monopole location $\vec y_a$. When the moduli $\vec y_a$ all coalesce, the two-cycles collapse to the same point and the space reduces to the ${\sf A}_{D-1}$ quotient singularity $\IR^4/\IZ_D$, where $D$ is the least common multiple of the integers $N$ and $K$. In the limit of dense monopole distribution, this reduction to four-dimensional flat space thus reinstates $x^4$ in the full eight-dimensional M-wave phase space, with the 3-algebra \eqref{eq:dualM3algebra}.


\section*{Acknowledgments}

We thank Christian S\"amann for helpful discussions. This work was supported in part by the COST Action MP1405 ``Quantum
Structure of Spacetime''. The work of DL and EM is supported by the
ERC Advanced Grant No.~320045 ``Strings and Gravity''. The work of RJS was
supported in part by the STFC Consolidated Grant ST/L000334/1
``Particle Theory at the Higgs Centre''. 

\bigskip

\bibliographystyle{JHEP}

\providecommand{\href}[2]{#2}\begingroup\raggedright\endgroup

\end{document}